\documentclass[twocolumn,showpacs,preprintnumbers,amsmath,amssymb]{revtex4}
\usepackage{graphicx}% Include figure files
\usepackage{dcolumn}% Align table columns on decimal point
\usepackage{bm}% bold math
\usepackage{epsfig}
\begin{document}

%\preprint{APS/123-QED}

\title{Comment on: A minimal model for beta relaxation in viscous liquids, by Jeppe C. Dyre and
Niels Boye Olsen, Phys. Rev. Lett. 91, 155703 (2003)}% Force line breaks with \\

\author{U. BUCHENAU}
 \email{u.buchenau@fz-juelich.de}
\author{S. KAHLE}
\affiliation{%
Institut f\"ur Festk\"orperforschung, Forschungszentrum J\"ulich\\
Postfach 1913, D--52425 J\"ulich, Federal Republic of Germany
}%

\date{October 23, 2003}% It is always \today, today,
             %  but any date may be explicitly specified
% PACS, the Physics and Astronomy
                             % Classification Scheme.

\maketitle

Dyre and Olsen \cite{dyre} report a key experiment for the
understanding of the glass transition, separating for the first
time the influence of the real temperature $T$ (the phonon bath
temperature) and the fictive temperature $T_f$ (the temperature
characterizing the thermodynamic state of the system) on the
Johari-Goldstein relaxation peak of a molecular glass former. They
find two surprising results: (i) an instantaneous increase of the
damping at the peak on heating, indicating a strong asymmetry
$\Delta$ of the potential minima of the relaxing units (ii) a
decrease of the peak frequency on the subsequent equilibration,
indicating an increase of the potential barrier between the minima
with increasing fictive temperature.

But one can also understand the experiment without these
counterintuitive assumptions of strong asymmetry and barrier
increase. A a recent model by one of us \cite{kvf} attributes the
damping to relaxing units distributed around the asymmetry zero
with a probability proportional to the Boltzmann factor
$\cosh(\Delta/2k_BT_f)$.

In order to obtain the barrier density $f_0(V)$ at the barrier
height $V$, one has to integrate over the asymmetry with the
weight factor $1/\cosh^2(\Delta/2k_BT)$. It is easy to convince
oneself that the above probability leads to $f_0(V)\sim T$ at
constant $T_f$, at least as long as $T$ is not too different from
$T_f$.

In addition, the model takes the elastic dipole interaction
between different relaxing entities into account. Combining eqs.
(6) and (7) of \cite{kvf}
\begin{equation}\label{fin}
f(V)=\frac{f_0(V)}{\left[1-3\int_0^Vf_0(v)dv\right]^{2/3}}.
\end{equation}
The enhancement of the measured barrier density $f(V)$ over the
true density $f_0(V)$ is due to all barriers lower than $V$. It
increases with increasing barrier height, and thus shifts the
Johari-Goldstein peak to higher barriers, as shown in Fig. 1. Both
the enhancement and the peak shift - this is the main point of the
Comment - increase with increasing relaxator density $f_0(V)$.

We model the experiment \cite{dyre} by a Johari-Goldstein peak in
$f_0(V)$. Since the damping is $\sim T f(V)$ and $f_0(V)$ is
proportional to temperature at constant $T_f$, the damping $\sim
T^{2+\delta}$, where $\delta$ comes from the enhancement factor in
eq. (\ref{fin}). From experiment, $\delta=0.6$. One fits this
requirement by the lorentzian in $f_0(V)$ shown in Fig. 1, thus
explaining the first surprising feature of the experiment.

The second surprising feature - the decrease of the relaxation
peak frequency on equilibration to a higher fictive temperature -
is explained by the increase of the damping, which shifts the peak
in $f(V)$ to higher barriers. As it turns out, $f_0(V)\sim
T_f^{7.9}$ describes both the long-time damping increase (Fig. 1
(b) of \cite{dyre}) and the long-time peak frequency shifts (Fig.
3 of \cite{dyre}).

This explanation does not require any unexpected properties of the
energy landscape. One only needs the theoretical concept of
independent relaxing entities, weakly coupled by the elastic
dipole interaction.

%%%%%%%%%%%%%%%%%%%%% begin figure %%%%%%%%%%%%%%%%%%%%%%%%%%%%%%%%%%%%%
\begin{figure}[b]
\hspace{-0cm} \vspace{0cm} \epsfig{file=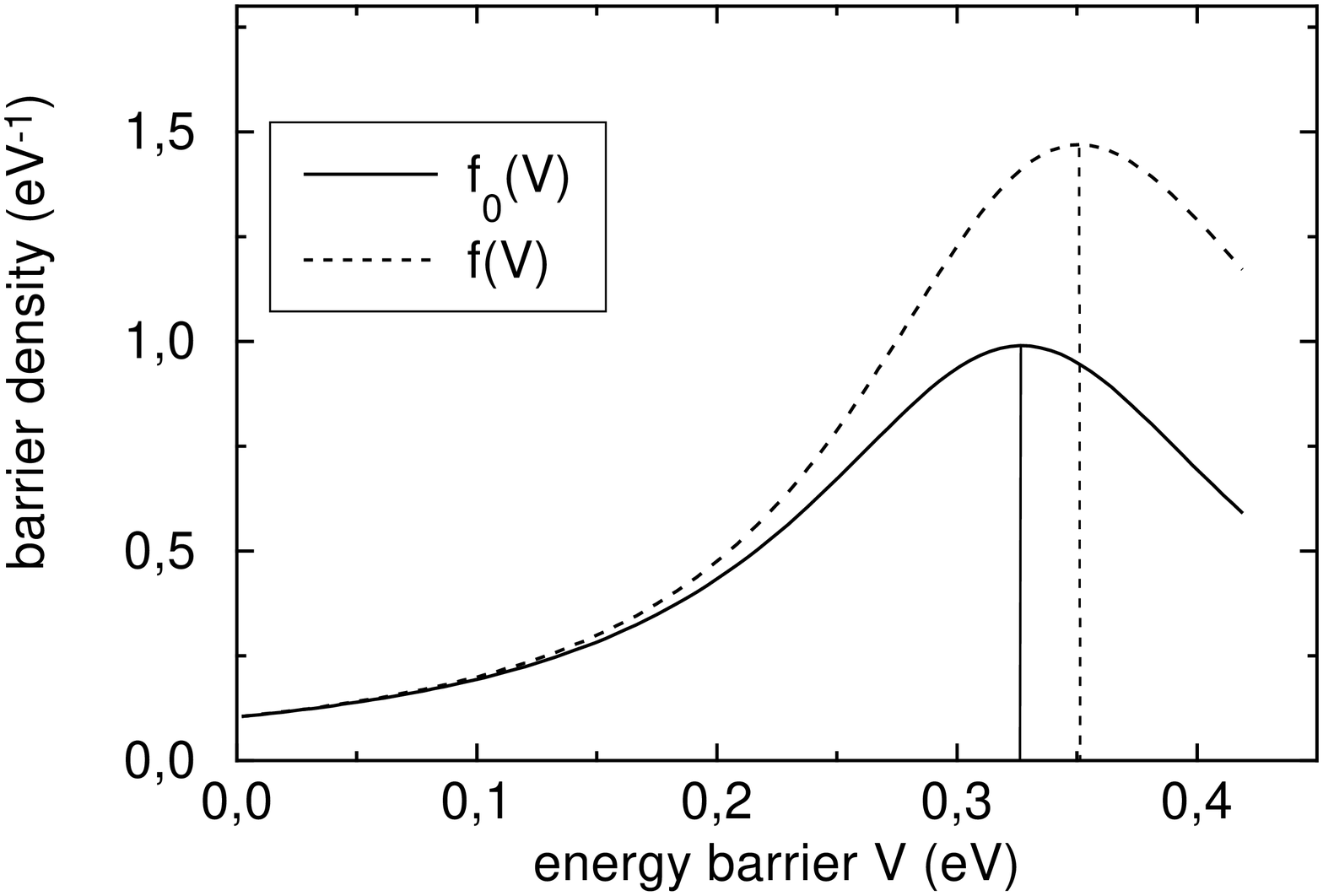,width=7
cm,angle=0} \vspace{0cm} \caption{Barrier densities in
tripropylene glycol at 184 K.}
\end{figure}
%%%%%%%%%%%%%%%%%%%%% end figure %%%%%%%%%%%%%%%%%%%%%%%%%%%%%%%%%%%%%%%

\end{document}